%% file: manuscript.tex
\documentclass[reprint, amsmath,amssymb,aps]{revtex4-2}

\usepackage{graphicx}
\usepackage{dcolumn}
\usepackage{bm}
\usepackage{amsmath}
\usepackage[usenames,dvipsnames]{xcolor}
\usepackage{mathtools}
\usepackage[colorlinks=true,citecolor=blue,linkcolor=blue]{hyperref}
\usepackage{bbm}
\usepackage{comment}
\usepackage{hyperref}
\usepackage{upgreek}
\usepackage{esint}
\usepackage{soul}
\usepackage[normalem]{ ulem } 
\usepackage{floatrow}
\usepackage{verbatim}
\usepackage{braket}
\usepackage{amsmath, amssymb}
\usepackage{array}
\usepackage{algorithm}
\usepackage{algpseudocode}
\usepackage{graphicx}
\usepackage{xcolor} 
\usepackage{enumitem}
\usepackage{subcaption}

\setlist[itemize]{parsep=1em, itemsep=1em} 
\setlist[enumerate]{parsep=0em, itemsep=1em} 

\newcommand{\PreserveBackslash}[1]{\let\temp=\\#1\let\\=\temp}
\newcolumntype{C}[1]{>{\PreserveBackslash\centering}p{#1}}

\newcommand{\rom}[1]{\uppercase\expandafter{\romannumeral #1\relax}}
\newcommand{\umqz}{U_\text{MQ}^\text{(Z)} }
\newcommand{\umqx}{U_\text{MQ}^\text{(X)} }
\newcommand{\umqp}{U_\text{MQ}^\text{(P)} }

\begin{document}

\title{Reduced constant-cost implementations of Clifford operations using global interactions}

\author{Jonathan Nemirovsky}
\author{Lee Peleg}
\author{Amit Ben Kish}
\author{Yotam Shapira}\email{yotam.shapira@quantum-art.tech}

\affiliation{Quantum Art, Ness Ziona 7403682, Israel}

\begin{abstract}
	We investigate quantum circuits built from arbitrary single-qubit operations combined with programmable all-to-all multiqubit entangling gates that are native to, among other systems, trapped-ion quantum computing platforms.
	We report a constant-cost of no more than 6 application of such Clifford entangling multiqubit gates to realize any sequence of Clifford operations of any length, without ancillae. 
	Furthermore, we show that any sequence of CNOT gates of any length, can be replaced with 5 applications of such Clifford entangling multiqubit gates, without ancillae. We investigate the required qubit drive power that is associated with these implementations. Our work introduces a practical and computationally efficient algorithm to realize these compilations.	
\end{abstract}

\maketitle

Clifford operations are central to many quantum information processing applications such as quantum error correction, simulation algorithms \cite{vandenberg2020circuit}, generation of pseudo-random unitaries \cite{haferkamp2023efficient,kueng2015qubit,zhu2017multiqubit}, compilation of quantum circuits and benchmarking \cite{elben2023randomized,huang2022foundations,huang2020predicting}. Thus, it is beneficial to obtain low-depth realizations of arbitrary Clifford operations.

Here we provide an explicit algorithm that decomposes arbitrary Clifford operations into a quantum circuit that utilizes single-qubit rotations and, at most, six multiqubit programmable all-to-all entangling gates, which is to the best of our knowledge a factor of $3$ better than the state of the art \cite{cleve2025improvedcliffordoperationsconstant}. Previous works have shown implementations of Clifford operations using such multiqubit gates with a gate count that is linear (in the number of qubit) \cite{maslov2018use,van2021constructing}, logarithmic \cite{maslov2022depth} or constant \cite{bravyi2022constant,cleve2025improvedcliffordoperationsconstant}, with Ref.~\cite{bravyi2022constant} greatly inspiring this work.

Explicitly, we consider multiqubit Clifford gates over $n$ qubits of the form,
\begin{equation}
	\umqp\left(\xi\right)=e^{-i\frac{\pi}{2}\sum_{k=1}^n \xi_{kk} P_k - i\frac{\pi}{4}\sum_{k>j}^n \xi_{kj} P_k P_j},
	\label{eqUmq}
\end{equation}
where $P_k$ is the $P\in\{X,Y,Z\}$ Pauli operator, acting on the $k$th qubit and $\xi\in\mathbb{F}_2^{n\times n}$ is an arbitrary symmetric matrix of zeros and ones, such that for $k \ne j$, $\frac{\pi}{4}\xi_{kj}$ is a correlated $P\otimes P$ rotation between the $k$th and $j$th qubits. We note that $\umqz$ is equivalent, up to single-qubit $Z$ rotations, to a layer of arbitrary Control-$Z$ ($\mathrm{CZ}$) gates. Similarly, $\umqx$, generates correlated $X\otimes X$ rotations and is equivalent up to single-qubit Hadamard gates to $\umqz$.

Gates of the form of Eq.~\eqref{eqUmq} arise naturally in trapped ions based quantum computers \cite{feng2023continuous,pogorelov2021compact,yao2022experimental,guo2024site,schwerdt2024scalable,nemirovsky2025efficientcompilationquantumcircuits}, where the long-range Coulomb coupling between the ions underpins the all-to-all programmability \cite{grzesiak2020efficient,shapira2020theory,shapira2023fast,lu2019global,bassler2023synthesis,lu2019global,lu2025implementing,shapira2025programmable}. We remark that additional quantum computing modalities have the potential to generate such programmable multiqubit interactions \cite{evered2023high,jeremy2021asymmetric,cooper2024graph,yongxin2025constant}, and can as well benefit from our method.

Similarly to Refs.~\cite{bravyi2022constant,bravyi2021hadamard}, we consider a decomposition of arbitrary Clifford operations to the form,
\begin{equation}
	U_\text{C}=-\text{L}-\text{CX}-\text{CZ}-\text{L}-\text{CZ}-\text{L}-\label{eqDecomp}
\end{equation} 
with $-\text{L}-$ a layer of single qubit gates, $-\text{CX}-$ a linear reversible circuit and $-\text{CZ}-$ a layer of Control-$Z$ gates. With this decomposition clearly two multiqubit gates are used for the two  $-\text{CZ}-$ layers, thus we focus on showing that the $-\text{CX}-$ layer can be implemented with at most five entangling multiqubit gates. Furthermore will show that one of the implementations of the $-\text{CX}-$ part ends with a $-\text{CZ}-$ layer that can be merged with the following $-\text{CZ}-$ layer of Eq. (\ref{eqSymMap}). Hence, the Clifford circuit is realized with at most six entangling multiqubit gates, compared to a lower bound of four multiqubit gates \cite{cleve2025improvedcliffordoperationsconstant}.

The linear reversible layer, $-\text{CX}-$, over $n$ qubits is represented by an invertible matrix, $M\in\mathbb{F}_2^{n\times n}$, such that the layer operates on a state in the computational basis as $\ket{\boldsymbol{v}}\mapsto\ket{M\boldsymbol{v}}$ (with addition performed modulo 2). Such an operation can be represented efficiently by a sequence of CNOT gate, e.g. by performing Gauss elimination of $M$, a circuit representing the layer can be formed with at most $n^2$ CNOTs. Moreover, methods to reduce this depth to $\mathcal{O}\left(n^2/\log\left(n\right)\right)$ are known in the literature \cite{patel2008optimal,patel302002efficient}. We assume such a CNOT circuit representing $-\text{CX}-$ is known (regardless of its depth or the way it is constructed).

Specifically, we make use of the symplecitc form  formalism, which we briefly recap. The symplectic form formalism provides an efficient representation of Clifford circuits, where each Clifford operator is mapped to a binary matrix acting on a vector space over the finite two-element field, $\mathbb{F}_2=\left\{0,1\right\}$~\cite{gottesman1997stabilizer,aaronson2004improved}.
Pauli vectors on $n$ qubits (also known as Pauli strings) are represented as binary vectors in 
$\mathbb{F}_2^{2n}$, defined by,
\begin{equation}
	(X_1^{a_1} Z_1^{b_1}) \otimes \cdots \otimes (X_n^{a_n} Z_n^{b_n}) \;\mapsto\; (a_1, \ldots, a_n \mid b_1, \ldots, b_n),\label{eqSymMap}
\end{equation}
where each pair $(a_i, b_i)$ specifies whether $X_i$, $Z_i$, or $Y_i$ acts on qubit $i$. Thus, the binary vector $(a \mid b)$ with $a,b \in \mathbb{F}_2^n$ encodes an $n$-qubit Pauli operator. A Clifford operator then acts linearly on these vectors via a symplectic matrix $S \in GL(2n, \mathbb{F}_2)$ representation, giving a group homomorphism from the Clifford group in $SU\left(2^n\right)$ to the symplectic group $Sp(2n, \mathbb{F}_2)\in\mathbb{F}_2^{2n\times2n}$. Importantly, up to an additional layer of single-qubit Pauli operators, Clifford operators admit a unique characterization through their symplectic matrix representation \cite{proctor2023simple}.

Pauli operators either commute or anti-commute. This is represented in their vector form by the matrix,
\begin{equation}
	\Omega = \begin{bmatrix}
		0 & -I_n \\
		I_n & 0
	\end{bmatrix},
\end{equation}
with $I_n$ the $n\times n$ identity operator, such that two Pauli vectors $u,v \in \mathbb{F}_2^{2n}$ commute (anti-commute) if $u^T \Omega v=0$ ($u^T \Omega v=1$) \cite{dehaene2003clifford}. Clifford operators must preserve these commutation relations, yielding the condition,
\begin{equation}
 	S^T\Omega S=\Omega.
 	\label{SymplecticConditon} 
\end{equation}
This ensures that the symplectic action faithfully captures the defining property of Clifford gates, i.e. mapping Pauli operators to Pauli operators while preserving commutation.

Let $E_{i,j}$ denote the $n \times n $ elementary matrix with a $1$ in position $(i,j)$ and 
0 in all other entries. When $i\ne j$, the two-qubit gates
$R_{Z_iZ_j}^{1/2} = e^{-i\frac{\pi }{4}Z_iZ_j}$ and $R_{X_iX_j}^{1/2}=e^{-i\frac{\pi}{4} X_iX_j}$ are represented by,

\begin{equation}	
 \small
	S(R_{Z_iZ_j}^{1/2}) =
	\begin{bmatrix}
		I_n & 0 \\
		E_{i,j} + E_{j,i} & I_n
	\end{bmatrix}, \; S(R_{X_iX_j}^{1/2}) =
	\begin{bmatrix}
		I_n & E_{i,j} + E_{j,i} \\
		0 & I_n
	\end{bmatrix}, \label{eqSzzSxx}
\end{equation}
respectively and for $i=j$ we have $R_{Z_i}^{1/2}=e^{-i\frac{\pi}{4}Z_i}$ and $R_{X_i}^{1/2}=e^{-i\frac{\pi}{4}X_i}$, represented by,
\begin{equation}
	S(R_{Z_i}^{1/2}) =
	\begin{bmatrix}
		I_n & 0 \\
		E_{i,i} & I_n
	\end{bmatrix}, \;\; S(R_{X_i}^{1/2}) =
	\begin{bmatrix}
		I_n & E_{i,i} \\
		0 & I_n
	\end{bmatrix}. \label{eqSzSx}
\end{equation}
Using the definition of $\umqz$ and $\umqx$ above and Eqs. (\ref{eqSzzSxx},\ref{eqSzSx}), 
we note that for symmetric $\xi\in\mathbb{F}_2^{n\times n}$ i.e.,
 $\xi_{n,m}=\xi_{m,n}\in\left\{0,1\right\}$, the multiqubit gate takes the form of a GCZ gate \cite{bravyi2022constant}. With this $\umqz(\xi)$ and $\umqx(\xi)$ are represented by,
\begin{equation}
	S(\umqz(\xi)) =
	\begin{bmatrix}
		I_n & 0 \\
		\xi & I_n
	\end{bmatrix}, \;\; S(\umqx(\xi)) =
	\begin{bmatrix}
		I_n & \xi \\
		0 & I_n
	\end{bmatrix},\label{eqSUMQ}
\end{equation}
respectively. 

Similarly, a CNOT gate with control qubit $i$ and target qubit $j$ is represented by the symplectic matrix,
\begin{equation}
	S_{\mathrm{CNOT}_{i \to j}} =
	\begin{bmatrix}
		I_n + E_{j,i} & 0 \\
		0 & I_n + E_{i,j}
	\end{bmatrix}.\label{eqSCNOT}
\end{equation}
Generalizing on this, the circuit representation of linear reversible, $-\text{CX}-$, layer takes the form,
\begin{equation}
	S_{\text{CX}} =
	\begin{bmatrix}
		A & 0 \\
		0 & B
	\end{bmatrix},\label{eqSCX}
\end{equation}
with $A,B\in\mathbb{F}_2^{n\times n}$ invertible matrices (due to the block form and the fact that $S$ is invertible). Furthermore, utilizing the symplectic condition in Eq.~\eqref{SymplecticConditon} and applying it on the block-diagonal symplectic form of Eq.~\eqref{eqSCX}, we obtain,
\begin{equation}
S_{\text{CX}}^T\Omega S_{\text{CX}}= \Omega.
\end{equation}
Recalling that $I_n=-I_n$ in $\mathbb{F}_2$ we conclude that,
\begin{equation} 
	B^T A = A^T B = I_n,
	\label{eq:symp_rul_ATB_is_I}
\end{equation}	
so that $A$ and $ B^T$ are reciprocals.

Next, we decompose $B=S_1 S_2$, with $S_k$ being symmetric matrices. Such a decomposition exists and is found efficiently \cite{taussky1972role}. With these identities and this decomposition we note that $S_{\text{CX}}$ is precisely formed by (see step-by-step realization in the SM \cite{SM}),
\begin{equation}
	\begin{split}
	S({\text{CX}})&=\\
	&S\left(\umqx\left({S_2^{-1}}\right)\right) S\left(\umqz\left({S_2}\right)\right) S\left(\umqx\left({S_2^{-1}+S_1}\right)\right)\\
	&S\left(\umqz\left({S_1^{-1}}\right)\right)S\left(\umqx\left({S_1}\right)\right),
	\end{split}\label{eqMagic}
\end{equation}
or alternatively,
\begin{equation}
	\begin{split}
		S({\text{CX}})&=\\
		&S\left(\umqz\left({S_2}\right)\right) S\left(\umqx\left({S^{-1}_2}\right)\right) S\left(\umqz\left({S_2+S^{-1}_1}\right)\right)\\
		&S\left(\umqx\left({S_1}\right)\right)S\left(\umqz\left({S_1^{-1}}\right)\right).
	\end{split}\label{eqMagic_alter}
\end{equation}
Finally, since the symplectic matrix determines the Clifford operator up to a layer of single qubit Pauli rotations, we conclude that the expression for the $-\text{CX}-$ layer comprises at most 5 entangling multiqubit gates and is given by,
\begin{equation}
	\begin{split}
		\text{CX}&= \umqx\left({S_2}\right)\umqz\left({S_2^{-1}}\right) \umqx\left({S_1+S_2^{-1}}\right)\\
		&\umqz\left({S_1^{-1}}\right)\umqx\left({S_1}\right)e^{i\frac{\pi}{2}{\sum_k\mu_kZ_k } } e^{i\frac{\pi}{2}{\sum_k\eta_kX_k }},
	\end{split}\label{eqMagic2}
\end{equation}
or alternatively,
\begin{equation}
	\begin{split}
		\text{CX}&= 
		\umqz\left({S_2^{-1}}\right)\umqx\left({S_2}\right)\umqz\left({S_1^{-1}+S_2}\right) \\
		&\umqx\left({S_1}\right)\umqz\left({S_1^{-1}}\right)e^{i\frac{\pi}{2}{\sum_k\tilde{\mu}_kZ_k } } e^{i\frac{\pi}{2}{\sum_k\tilde{\eta}_kX_k }},
	\end{split}\label{eqMagic2_alter}
\end{equation}
where the coefficients $\mu_k,\eta_k,\tilde{\mu}_k,\tilde{\eta}_k \in \{0,1\}$ are determined by the symplectic phase vector associated with the  $-\text{CX}-$ layer and the gates on the right hand-side of Eq.~\eqref{eqMagic2} and Eq.~\eqref{eqMagic2_alter}.

Crucially, the latter decomposition, in Eq.~\eqref{eqMagic2}, terminates with a $\umqz$ gate, and can therefore be merged with the following $-\text{CZ}-$ layer, further reducing the multiqubit gate count.

We note that in the case that $B$ from Eq.~\eqref{eqSCX} is symmetric then the decomposition $B=S_1 S_2$ is trivially satisfied with $S_1=I$, resulting in at most three multiqubit gates in Eqs.~\eqref{eqMagic2} and \eqref{eqMagic2_alter}.

Next, we investigate the drive power necessary for the implementation of our decomposition. The drive power is an important quantity as many fundamental gate-errors naturally scale with the drive power, e.g. unwanted photon scattering. For trapped-ions qubits, it has been shown that the gate drive power scales as $\text{nuc}\left(\xi\right)$, the nuclear-norm of $\xi$, i.e. the $L_1$ norm of its eigenvalues \cite{shapira2023fast}. To investigate this aspect of our decomposition we generate random linear reversible layers, $M$, and construct their corresponding CNOT circuit with Gaussian elimination. We then decompose them according to Eqs.~\eqref{eqMagic} and \eqref{eqMagic_alter}. 

There are many degrees of freedom in the decomposition of $B$ to a product of two symmetric matrices. These are utilized in order to reduce, in a graph walker method, the resulting total nuclear norm of the decomposition, $\Omega_\text{nuc}$. Addition reduction is obtained by virtual qubit permutations, as done in Refs.~\cite{nemirovsky2025efficientcompilationquantumcircuits,baldwin2022re}. 

\begin{figure}
	\centering
	\includegraphics[width=0.8\textwidth]{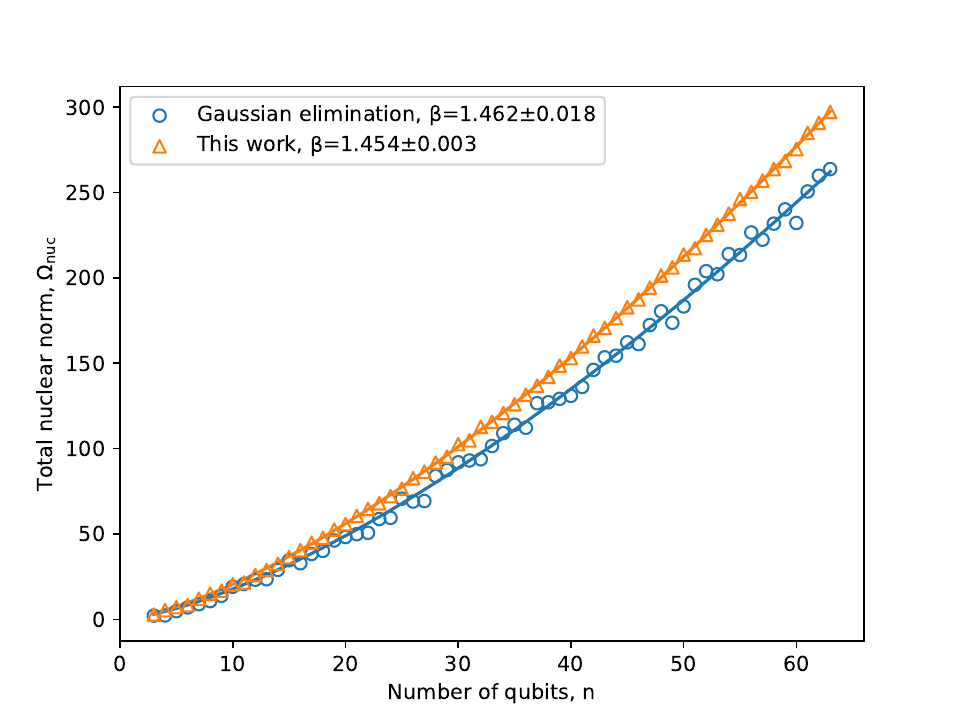} 
	\caption{Total nuclear norm of an implementation using a standard Gaussian elimination to $\sim n^2$ two-qubit gate followed by a merge of parallel operations (blue circles), compared to the total nuclear norm of the constant-cost implementation of this work (orange triangles), showing that the depth reduction is performed with comparable resources. We fit this data to a power law (solid), showing that the total nuclear norm scales approximately as $\sim n^{3/2}$ for both methods.}
	\label{fig:nuclear}
\end{figure}

Figure \ref{fig:nuclear} shows the total nuclear norm of the decomposition, $\Omega_\text{nuc}$, of random instances of $-\text{CX}-$ (orange triangles) for varying number of qubits between 3 and 63 (horizontal). This is compared to the total nuclear norm used by a standard decomposition based on the Gaussian elimination, i.e. obtained by implementing CNOT gates with two-qubit $Z\otimes Z$ gates and single-qubit gates and merging parallel operations to MQ gates (blue circles). As seen, our constant-cost implementation has a similar total nuclear norm to that of the naive implementation. We fit these results to a power law (solid), $\Omega_\text{nuc}\propto n^\beta$, with the results appearing in the legend, showing a consistent scaling within fit error of less than $n^{3/2}$ for both methods.

In conclusions, we have shown a computationally efficient algorithm that implements Clifford operation at a constant-cost of 6 multiqubit gates, based on an implementation of linear reversible, CNOT, circuits with 5 multiqubit gates. We have further investigated the drive power associated with this implementation, showing that it can be achieved with the same resources as standard Gaussian elimination methods, that is, depth and execution times are reduced without incurring additional power. 

Beyond immediate applications of these decomposition, we note that it can work in conjunction with generic compilation techniques, where the circuit to be compiled can be `cut' to blocks separated by Clifford operations, that are then realized with a constant-cost.

\bibliographystyle{unsrt}
\bibliography{references}

\cleardoublepage
\onecolumngrid
{
	\begin{center}
	{\large \bfseries Supplemental material \par}
	\end{center}	
	\bigskip
	\setcounter{section}{0}
	{\input{SM.tex}}
	\unskip
	}

\end{document}

%% file: SM.tex

\section{Step-by-step derivation of Eq. (\ref{eqMagic}) and (\ref{eqMagic_alter})}
The symplectic matrix of a general CNOT circuit operating on $n$ qubits has the form,
\begin{equation}
	S\left(\prod_{k=1,2,3,...,n} {\mathrm CNOT_{i_k\rightarrowtail j_k}}  \right) =
	\begin{bmatrix}
		A & 0 \\
		0 & B
	\end{bmatrix}
\end{equation}
with $A$ and $B$ being $n\times n$ matrices over the field $\mathbb{F}_2$, satisfying Eq. (\ref{eq:symp_rul_ATB_is_I}), i.e.,
\begin{equation}
A^T B = B^T  A = I.
\label{eq:symp_cond_AB}
\end{equation}
For every $B\in GL(n,\mathbb{F}_2)$ one can always find symmetric matrices $S_1, S_2 \in  GL(n,\mathbb{F}_2)$ such that 
$ B=S_2 S_1 $ \cite{taussky1972role}. From Eq. (\ref{eq:symp_cond_AB}) we conclude that 

\begin{equation}
	B^T A = S_1 S_2 A = I.
	\label{eq:SA_rule}
\end{equation} 

We now perform the following steps,
\begin{enumerate}
	\item We decompose $B$ to a product of two symmetric matrices $S_1, S_2 \in  GL(n,\mathbb{F}_2)$ so that,
		\[ 	
		\begin{bmatrix}
			A & 0 \\
			0 & B
		\end{bmatrix} = 
		\begin{bmatrix}
			A & 0 \\
			0 & S_2 S_1
		\end{bmatrix}. 
		 \]
	\item Since $S_2$ is invertible we can apply a,
	$S(\umqx\left(S_2^{-1}\right))=\begin{bmatrix}
		I & S^{-1}_2 \\
		0 & I
	\end{bmatrix}. $
	\[S(\umqx\left(S_2^{-1}\right))
	\begin{bmatrix}
	A & 0 \\
	0 & S_2 S_1
	\end{bmatrix}=	\begin{bmatrix}
	A & S_1 \\
	0 & S_2 S_1
	\end{bmatrix}.
	 \]
	\item 	We then apply a $S(\umqz\left(S_2)\right))=\begin{bmatrix}
		I & 0 \\
		S_2 & I
	\end{bmatrix}$ gate so that with Eq. (\ref{eq:SA_rule}) we obtain,
		\[S(\umqz\left(S_2\right))
	\begin{bmatrix}
		A & S_1 \\
		0 & S_2 S_1
	\end{bmatrix}=
	\begin{bmatrix}
		A & S_1 \\
		S_2 A & S_2 S_1 + S_2 S_1
	\end{bmatrix}=
	\begin{bmatrix}
		A & S_1 \\
		S_2 A & 0
	\end{bmatrix}=
	\begin{bmatrix}
	A & S_1 \\
	S_1^{-1} & 0
	\end{bmatrix}.
	\]
	\item 	We then apply a  $S(\umqx\left(S_2^{-1})\right))=\begin{bmatrix} 
		I & S_2^{-1} \\
		0 & I
	\end{bmatrix}$ and again with Eq. (\ref{eq:SA_rule}) we obtain,
	\[S(\umqx\left(S_2^{-1}\right))
	\begin{bmatrix}
		A & S_1 \\
		S_1^{-1} & 0
	\end{bmatrix}=
	\begin{bmatrix}
		A+S_2^{-1}S_1^{-1} & S_1 \\
		S_1^{-1} & 0
	\end{bmatrix}=
	\begin{bmatrix}
	A+A & S_1 \\
	S_1^{-1} & 0
	\end{bmatrix}=
	\begin{bmatrix}
		0 & S_1 \\
		S_1^{-1} & 0
	\end{bmatrix}.
	\]
	\item 	We then apply a  $S(\umqx\left(S_1)\right))=\begin{bmatrix} 
	I & S_1 \\
	0 & I
	\end{bmatrix}$ and obtain,
	\[S(\umqx\left(S_1\right))
	\begin{bmatrix}
		0 & S_1 \\
		S_1^{-1} & 0
	\end{bmatrix}=
	\begin{bmatrix}
		I & S_1 \\
		S_1^{-1} & 0
	\end{bmatrix}.
	\]
	\item 	And then apply a $S(\umqz\left(S_1^{-1})\right))=\begin{bmatrix} 
	I & 0 \\
	S_1^{-1} & I
	\end{bmatrix}$ which leads to,
	\[S(\umqz\left(S_1^{-1}\right))
	\begin{bmatrix}
		I & S_1 \\
		S_1^{-1} & 0
	\end{bmatrix}=
	\begin{bmatrix}
		I & S_1 \\
		0 & I
	\end{bmatrix}.
	\]	
\end{enumerate}
The right-hand side of the last equation is by definition $S\left(\umqx\left(S_1\right)\right)$. Thus we conclude that, if 
$S(\text{CNOT})=
\begin{bmatrix}
	A & 0 \\
	0 & B
\end{bmatrix} = 
\begin{bmatrix}
	A & 0 \\
	0 & S_2 S_1
\end{bmatrix}$ then,
\[
	S\left(\umqz\left(S_1^{-1})\right)\right) \cdot S\left(\umqx\left(S_1)\right)\right)\cdot S\left(\umqx\left(S_2^{-1})\right)\right) \cdot S\left(\umqz\left(S_2)\right)\right)\cdot  S\left(\umqx\left(S_2^{-1}\right)\right) \cdot S\left(\text{CNOT}\right) = S\left(\umqx\left(S_1)\right)\right)
\]
and recalling that, 
\[S\left(\umqx\left(S')\right)\right)S\left(\umqx\left(S'')\right)\right)=S\left(\umqx\left(S'+S'')\right)\right)\] and \[S\left(\umqz\left(S')\right)\right)S\left(\umqz\left(S'')\right)\right)=S\left(\umqz\left(S'+S'')\right)\right)\] we conclude that,
\begin{equation}
	S(\text{CNOT})= S\left(\umqx\left(S_2^{-1}\right)\right)S\left(\umqz\left(S_2)\right)\right)
	S\left(\umqx\left(S_2^{-1}+S_1)\right)\right)
	S\left(\umqz\left(S_1^{-1})\right)\right)S\left(\umqx\left(S_1)\right)\right).
\end{equation}
Where we used the fact that $S\left(\umqz\right)$ is its own inverse under $\mathbb{F}_2$, or equivalently that $\umqz$ is its own inverse up to single-qubit Pauli operators. This completes the proof of Eq. (\ref{eqMagic}).

Eq. (\ref{eqMagic_alter}) is proved in a similar manner by multiplying from the right (instead of from the left) the symplectic matrix 
$S(\text{CNOT})=\begin{bmatrix}
	A & 0 \\
	0 & S_2 S_1
\end{bmatrix} = 
\begin{bmatrix}
S_2^{-1} S_1^{-1} & 0 \\
0 & S_2 S_1
\end{bmatrix} 
$ with appropriate symplectic matrices.
To see this explicitly we perform the following steps:
\begin{enumerate}
	\item Decomposing $B$ to a product of two symmetric matrices $S_1, S_2 \in  GL(n,\mathbb{F}_2)$ so that,
	\[ 	
	\begin{bmatrix}
		A & 0 \\
		0 & B
	\end{bmatrix} = 
	\begin{bmatrix}
		A & 0 \\
		0 & S_2 S_1
	\end{bmatrix}. 
	\]
	\item Since $S_2$ is invertible we can apply a,
	$S(\umqz\left(S_1^{-1}\right))=\begin{bmatrix}
		I & 0 \\
		S^{-1}_1 & I
	\end{bmatrix} $ from the right hand-side. So that,
	\[
	\begin{bmatrix}
		A & 0 \\
		0 & S_2 S_1
	\end{bmatrix}S(\umqz\left(S_1^{-1}\right))=	\begin{bmatrix}
		A & 0 \\
		S_2 & S_2 S_1
	\end{bmatrix}.
	\]
	\item 	We then apply a $S(\umqx\left(S_1)\right))=\begin{bmatrix}
		I & S_1 \\
		0 & I
	\end{bmatrix}$ from the right hand-side so that with Eq. (\ref{eq:SA_rule}) we obtain,
	\[
	\begin{bmatrix}
		A & 0 \\
		S_2 & S_2 S_1
	\end{bmatrix}S(\umqx\left(S_1\right))=
	\begin{bmatrix}
		A & A S_1 \\
		S_2  & S_2 S_1 + S_2 S_1
	\end{bmatrix}=
	\begin{bmatrix}
		A & A S_1 \\
		S_2 & 0
	\end{bmatrix}=
	\begin{bmatrix}
		A & S_2^{-1} \\
		S_2 & 0
	\end{bmatrix}.
	\]
	\item 	We then apply a  $S(\umqz\left(S_1^{-1})\right))=\begin{bmatrix} 
		I & 0 \\
		S_1^{-1} & I
	\end{bmatrix}$ from the right side and again with Eq. (\ref{eq:SA_rule}) we obtain,
	\[
	\begin{bmatrix}
		A & S_2^{-1} \\
		S_2 & 0
	\end{bmatrix}S(\umqz\left(S_1^{-1}\right))=
	\begin{bmatrix}
		A+S_2^{-1}S_1^{-1} & S_2^{-1} \\
		S_2 & 0
	\end{bmatrix}=
	\begin{bmatrix}
		A+A & S_2^{-1} \\
		S_2 & 0
	\end{bmatrix}=
	\begin{bmatrix}
		0 & S_2^{-1} \\
		S_2 & 0
	\end{bmatrix}.
	\]
	\item 	We then apply a  $S(\umqz\left(S_2)\right))=\begin{bmatrix} 
		I & 0 \\
		S_2 & I
	\end{bmatrix}$ from the right and obtain,
	\[
	\begin{bmatrix}
		0 & S_2^{-1} \\
		S_2 & 0
	\end{bmatrix}S(\umqz\left(S_2\right))=
	\begin{bmatrix}
		I & S_2^{-1} \\
		S_2 & 0
	\end{bmatrix}.
	\]
	\item 	And then apply a $S(\umqx\left(S_2^{-1})\right))=\begin{bmatrix} 
		I & S_2^{-1} \\
		0 & I
	\end{bmatrix}$ from the right, which leads to,
	\[
	\begin{bmatrix}
		I & S_2^{-1} \\
		S_2 & 0
	\end{bmatrix}S(\umqx\left(S_2^{-1}\right))=
	\begin{bmatrix}
		I & 0 \\
		S_2 & I
	\end{bmatrix}.
	\]	
\end{enumerate}
The right-hand side of the last equation is by definition $S\left(\umqz\left(S_2\right)\right)$. Thus we conclude that, if 
$S(\text{CNOT})=
\begin{bmatrix}
	A & 0 \\
	0 & B
\end{bmatrix} = 
\begin{bmatrix}
	A & 0 \\
	0 & S_2 S_1
\end{bmatrix}$ then,
\[
\left(\text{CNOT}\right) 
\cdot
S\left(\umqz\left(S_1^{-1}\right)\right) 
\cdot
S\left(\umqx\left(S_1)\right)\right) 
\cdot
S\left(\umqz\left(S_1^{-1})\right)\right) 
\cdot
S\left(\umqz\left(S_2)\right)\right)
\cdot
S\left(\umqx\left(S_2^{-1})\right)\right)
= S\left(\umqz\left(S_2)\right)\right).
\]
So that finally we conclude that,
\begin{equation}
	S(\text{CNOT})= S\left(\umqz\left(S_2\right)\right)S\left(\umqx\left(S_2^{-1})\right)\right)
	S\left(\umqz\left(S_2+S_1^{-1})\right)\right)
	S\left(\umqx\left(S_1)\right)\right)S\left(\umqz\left(S_1^{-1})\right)\right).
\end{equation}